%% file: A.tex
\documentclass[12pt,a4paper,fleqn]{article} 
\usepackage{tcolorbox}
\usepackage[bottom]{footmisc} 
\usepackage[fleqn]{amsmath} 
\usepackage{amssymb} 
\usepackage{physics} 
\usepackage{graphicx} 
\usepackage{cite} 
\usepackage[pdftex]{hyperref} 
\usepackage{ifthen}
\usepackage{bm}
\usepackage{enumitem}
\input{defs.tex} 

\begin{document} 

\title{ 
\textbf{Statistical analysis of event classification \\
in  experimental data} 
}
\author{
\textit{Rudolf Fr\"uhwirth}\footnote{ 
\quad \href{mailto:rudolf.fruehwirth@oeaw.ac.at}
{\texttt{mailto:rudolf.fruehwirth@oeaw.ac.at} } } 
\ and \textit{Winfried Mitaroff}
\\
Institute of High Energy Physics, \\
Austrian Academy of Sciences, Vienna 
}
\date{28 June 2023}   

\maketitle

\begin{abstract} 

The paper addresses general aspects of experimental data analysis, 
dealing with the separation of ``signal vs. background''. 
It consists of two parts. 

Part  I is a tutorial on statistical event classification, 
Bayesian inference, and test optimization. 
Aspects of the base data sample if 
being created by Poisson processes are discussed, 
and a method for estimating the 
unknown numbers of signal and background events is presented. 
Data quality of the selected events sample is assessed by 
the expected purity and background contamination. 

Part II contains a rigorous statistical analysis of the methods discussed in Part~I.
Both Bayesian and frequentist estimators of the unknown signal/background content are investigated. The estimates and their stochastic uncertainties are calculated for various conjugate priors in the Bayesian case, and for three choices of the virtual parent population in the frequentist case.
\end{abstract} 


\newpage 


\section{Introduction} 
\label{sec:intro} 

The analysis of experimental data 
-- e.g. events created at a particle collider and recorded in a detector -- 
often requires selecting those rare events with a signature of interest (the ``signal'') 
and rejecting the bulk of ``background''. 
This is performed by some ``test'', modelled according to the signature, 
in order to make a decision (\textit{positive} or \textit{negative}) of whether 
the hypothesis ``event is signal'' is \textit{true} or \textit{false}. 

In practice, such a test is never perfect: some signal events may wrongly be rejected 
(error type~I), and some background may wrongly be selected (error type~II). 
It is therefore essential to know the fraction of ``wrong events'' 
in the selected and rejected data samples, 
which are characteristic values of the test being used. 

These values are not known \aprior, but may be obtained by applying 
the very same test on two samples of Monte Carlo generated data, 
simulating as accurate as possible pure signal and pure background, respectively. 
The test's ``characteristic values'' (error types~I and II and their complements: 
sensitivity and specificity) are conditional probabilities, 
represented by the $2 \times 2$ ``contingency matrix''. 

The test criteria depend on parameters 
which influence the selection and rejection characteristics. 
Optimization may be achieved by defining an appropriate ``figure of merit'', 
to be maximized by varying the test parameters. 

The conditional posterior probabilitiy of a selected event being genuine signal, 
provided the test result is positive, can be calculated by Bayesian inference. 
It depends on the prior probability (prevalence) which is in general unknown, however, 
can be estimated by using Monte Carlo information together with real data. 

Part I presents a tutorial on the basics (Section~\ref{sec:basix}), 
the signal and background simulations for obtaining the test charcteristics 
as encoded in the contingency matrix (Section~\ref{sec:simul}), 
and their use in Bayesian inference (Section~\ref{sec:bayes}). 
The scenario of the base sample being created 
by two independent Poisson processes for 
signal and background is regarded in Section~\ref{sec:opti}. 
Postulating these numbers to be fixed, 
estimates for the number of selected and rejected events are given 
(Subsection~\ref{sec:opti-1}); test optimization 
and the choice of a ``figure of merit'' 
are discussed in Subsection~\ref{sec:opti-3}. 
Section~\ref{sec:prior} presents a method for estimating the 
unknown numbers of signal and background events 
from the known numbers of selected and rejected events, 
using Monte Carlo information; 
estimates of the Bayesian prior probabilities follow straight-forward. 
Section~\ref{sec:quali} shows how to assess the quality of selected data 
by its purity or its fraction of background contamination. 

Part II follows up with a comprehensive and rigorous statistical analysis. 
Section~\ref{sec:Assumptions} recapitulates the basic assumptions and the notation. 
Bayesian estimation is the topic of Section~\ref{sec:BayesEst}. The posterior distribution of the fraction of events classified as signal is computed for various conjugate priors, both ``uninformative'' and informative. A simple affine transformation yields the posterior mean and the posterior variance of the unknown signal content of the sample.  Frequentist estimation, on the other hand, presupposes a virtual parent population from which the observed sample is drawn (Section~\ref{sec:FrequEst}). The frequentist estimators and their covariance matrices are calculated for three choices of the parent population:  fixed sample size (Subsection~\ref{subsec:fixed}), Poisson-distributed sample size (Subsection~\ref{subsec:Poisson}), and sample size drawn from a mixture of Poisson distribution (Subsection~\ref{subsec:mixture}). The pros and cons of the two approaches are summarized in the final Section~\ref{sec:Discussion}.





\vspace*{\baselineskip}
\begin{center} 
{\Large \textbf{Part I: The tutorial} } 
\end{center} 

\section{The basics} 
\label{sec:basix} 

Given a finite data sample $\setD$ of $N$ random events,\footnote{ 
\, Aspects of $\setD$ if having emerged from two Poisson processes 
are in short discussed in Section \ref{sec:opti}, 
and a thorough analysis is presented in part II of this paper.} 
where each element $\evt_i$ is theoretically classified as being either a 
``signal event'' 
belonging to sub-sample $\setS$, 
or a ``background event'' 
belonging to sub-sample $\setB = \setD \setminus \setS$, 
with the number of events 
being $N_S$ and $N_B$, respectively. Then, trivially 
\begin{equation} \label{eq:x-1} 
N_S + N_B = N. 
\end{equation}
\indent 
For a particular element $\evt_i$, define the ``null hypothesis'' 
$\mathcal{H} \equiv \evt_i \in \setS$ as the event being signal, and 
the ``alternative hypothesis'' $\bar{\mathcal{H}} \equiv \evt_i \in \setB$ 
as it being background. 
The probabilities (in general unknown) of $\evt_i$ being either  ``signal'' 
($\mathcal{H}$ = \textit{true}) 
or ``background'' ($\mathcal{H}$ = \textit{false}) are the fractions 
\begin{equation} \label{eq:x-2} 
p_S \equiv\Pr \, (\mathcal{H}) = \frac{N_S}{N} \le 1 
\; \ldots \; \mathit{prevalence}, \qquad 
p_B \equiv\Pr \, (\bar{\mathcal{H}}) = \frac{N_B}{N} = 1 - p_S,
\end{equation}
with their ratio being 
\begin{equation} \label{eq:x-3} 
R_{S/B} = \frac{N_S}{N_B} = \frac{p_S}{p_B} 
\; \ldots \; \mathit{signal \, to \, background \, ratio} \, \mathrm{(SBR).} 
\end{equation}
\indent 
If the SBR is \aprior{} known (either from theory or from previous experiments) then the 
prior probabilities are trivially 
\begin{equation} \label{eq:x-4} 
p_S = \frac{R_{S/B}}{\, 1 + R_{S/B} \,} \quad\mathrm{and} \quad 
p_B = \frac{1}{\, 1 + R_{S/B} \,} \, ,
\end{equation} 
but often the SBR may only be guessed by order of magnitude. 
In case of the signal consisting of \textit{rare events}, $R_{S/B} \ll 1$. 

It is assumed that some \textit{test} is available which aims at determining, 
for each element $\evt_i \in \setD$, 
whether it does comply with hypothesis $\mathcal{H}$ and therefore will be 
\textit{selected} and put into sub-sample $\setSa$; 
otherwise it will be \textit{rejected} and put into 
sub-sample $\setBa = \setD \setminus \setSa$. 
The ``positive test result'' $\mathcal{T} \equiv \evt_i \in \setSa$ and 
the ``negative test result'' $\bar{\mathcal{T}} \equiv \evt_i \in \setBa$ 
are defined accordingly. 

However, any test is inherently not perfect but prone to errors of type I and II, the rates of 
which must be evaluated. 
 

\section{MC simulation} 
\label{sec:simul} 
 
In order  to achieve this, two separate samples are created by 
\textit{Monte Carlo simulation}: $\setSp$ containing $N'_S$ pure signal events, 
and $\setBp$ containing $N'_B$ pure background events. 
Their ratio may be chosen arbitrarily; often $N'_S / N'_B \gg R_{S/B}$. 


The \textit{simulated to real background ratio} is 
\begin{equation} \label{eq:s-0} 
r_B = \frac{N'_B}{N_B} = \frac{N'_B}{N} \cdot \left( 1 + R_{S/B} \right) 
\approx \frac{N'_B}{N} \quad (\mathrm{approximation \ if} \: R_{S/B} \ll 1), 
\end{equation} 
and its inverse $1 / r_B$ may serve as the \textit{normalization factor} for scaling 
simulated to real background, as used in Section \ref{sec:quali}. 

The test criteria are described by a set of parameters $\vec{\theta}$. 
Applying the test on the $\setSp$ signal and $\setBp$ background MC samples yield 
\footnote{ 
\, Deliberately using the terms ``right'' and ``wrong'' instead of ``true'' and ``false''.} 
\begin{itemize}[itemsep=2pt]
\item[] 
$n_{S+}$ = no. of signal MC events selected = \textit{right positives}, 
 \item[] 
$n_{S-}$ = no. of signal MC events rejected =  \textit{wrong negatives}, 
 \item[] 
$n_{B+}$ = no. of background MC events selected =  \textit{wrong positives}, 
 \item[] 
$n_{B-}$ = no. of background MC events rejected =  \textit{right negatives}, 
\end{itemize} 
\begin{equation} \label{eq:s-1} 
n_{S+} + n_{S-} = N'_S, \qquad \qquad n_{B+} + n_{B-} = N'_B,
\end{equation} 
\vspace{-7mm} 
\begin{equation}  \label{eq:s-2} 
n_{S+} + n_{B+} = N'_{+}, \qquad \qquad n_{S-} + n_{B-} = N'_{-} ,
\end{equation} 
defining the conditional probabilities 
\footnote{ 
\, The term \textit{purity}, which is often defined ambigously, 
will be used in Section \ref{sec:quali}.} 
\begin{eqnarray} \label{eq:s-3} 
\varepsilon &=& \frac{n_{S+}}{N'_S} \, = \,\Pr \, (\mathcal{T} \, | \, \mathcal{H}) 
\; \ldots \; \mathit{efficiency, sensitivity, select | hit \, rate}, 
\\ [2mm] \label{eq:s-4} 
\alpha &=& \frac{n_{S-}}{N'_S} \, = \,\Pr \, (\bar{\mathcal{T}} \, | \, \mathcal{H}) 
\; \ldots \; \mathit{error \, type \, I, significance, loss | miss \, rate}, 
\\ [2mm] \label{eq:s-5} 
\beta &=& \frac{n_{B+}}{N'_B} \, = \,\Pr \, (\mathcal{T} \, | \, \bar{\mathcal{H}}) 
\; \ldots \; \mathit{error \, type \, II, contamination | fake \, rate}, 
\\ [2mm] \label{eq:s-6} 
\eta &=& \frac{n_{B-}}{N'_B} \, = \,\Pr \, (\bar{\mathcal{T}} \, | \, \bar{\mathcal{H}}) 
\; \ldots \; \mathit{test \, power, specificity, reject \, rate}, \quad \mathrm{and} 
\\ [2mm] \label{eq:s-7} 
\ell_{+} &=& \frac{\varepsilon}{\beta} 
\; \ldots \; \mathit{positive \, likelihood \, ratio}, 
\\ [2mm] \label{eq:s-8} 
\ell_{-} &=& \frac{\alpha}{\eta} 
\; \ldots \; \mathit{negative \, likelihood \, ratio}, 
\end{eqnarray} 
\vspace{-5 mm} 
\begin{equation} \label{eq:s-9} 
\mathrm{trivially \; obeying} \qquad 
\varepsilon + \alpha = 1, \quad \beta + \eta = 1 \quad \Longrightarrow \quad 
\varepsilon \, \eta - \alpha \, \beta = \varepsilon - \beta .
\end{equation} 
\indent 
The Monte Carlo samples $\setSp$ and $\setBp$ contain randomly simulated events, 
hence the test's characteristic values $\left[ \, \varepsilon, \alpha, \beta, \eta \, \right]$  
in Eqs.~(\ref{eq:s-3}--\ref{eq:s-6}) are stochastic and represent empirical means. 
Their statistical errors 
$\sigma (\varepsilon), \, \sigma (\alpha) \propto 1 / \sqrt{N'_S}$ and 
$\sigma (\beta), \, \sigma (\eta) \propto 1 / \sqrt{N'_B}$ scale down 
with the sample sizes $N'_S$ and $N'_B$ increasing, 
thus may be neglected for big enough samples. 
However, improper simulation of reality can introduce systematic errors 
which are difficult to detect and may bias subsequent results. 

The characteristic values (\ref{eq:s-3}--\ref{eq:s-6}) resulting from above test 
may be arranged in the $2 \times 2$ 
\textit{contingency/confusion/error matrix} 
\footnote{ 
\, Note: in the literature these names are sometimes used for the transposed matrix 
$\cMat^T$ instead.} 
\begin{equation} \label{eq:s-10} 
\begin{array}{l} 
\cMat \; = \; 
\left( \begin{array}{cc} 
\varepsilon = \,\Pr \, (\mathcal{T} \, | \, \mathcal{H}) & \qquad 
\beta = \,\Pr \, (\mathcal{T} \, | \, \bar{\mathcal{H}}) \\ [2mm]
\alpha = \,\Pr \, (\bar{\mathcal{T}} \, | \, \mathcal{H}) & \qquad 
\eta = \,\Pr \, (\bar{\mathcal{T}} \, | \, \bar{\mathcal{H}}) 
\end{array} \right), 
\qquad \quad \mbox{with} 
\\[8mm]
\left( \begin{array}{c} 
N'_{+} \\[2mm] 
N'_{-}
\end{array} \right) \; = \; \cMat \cdot 
\left( \begin{array}{c} 
N'_S \\[2mm] 
N'_B 
\end{array} \right) 
\qquad \quad \mbox{and} \qquad \quad 
\det \, ( \cMat ) = \delta = \varepsilon - \beta .
\end{array} 
\end{equation} 
\indent
A meaningful test should be designed such that 
\begin{equation} \label{eq:s-11} 
\alpha + \beta < 1 \quad \Longrightarrow \quad \delta = \varepsilon - \beta > 0 
\quad \Longrightarrow \quad \ell_{+} = \varepsilon / \beta > 1, \quad \ell_{-} = \alpha / \eta < 1.
\end{equation} 


\section{Bayesian inference}
\label{sec:bayes} 

\begin{figure}[h!t] 
\centering 
\includegraphics*[width=0.75\textwidth]{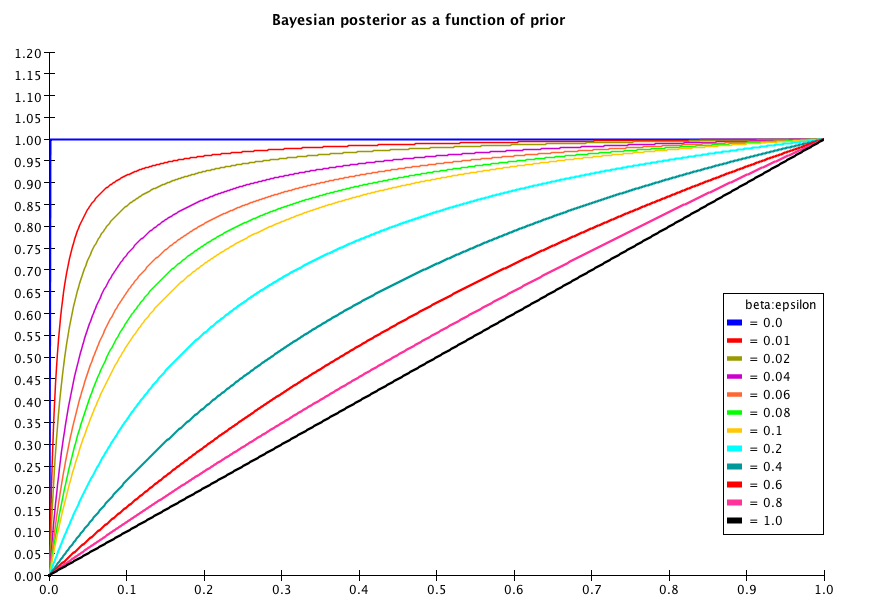} 
\caption[]{ 
Bayesian posterior $p_{S+}$ as a function of $p_S$ for different 
$1 / \ell_{+} = \beta / \varepsilon$. }
\label{fig:bayes} 
\end{figure}

The values $\left[ \, \varepsilon, \alpha, \beta, \eta \, \right]$, 
defined in (\ref{eq:s-3}--\ref{eq:s-6}), represent the conditional probabilities of 
the test being \textit{positive} or \textit{negative}, provided 
the hypothesis $\mathcal{H}$ is \textit{true} or \textit{false}. 
Asking inversely: which are the conditional posterior probabilities 
of hypothesis $\mathcal{H}$ being \textit{true} or \textit{false}, provided 
this test has been \textit{positive} or \textit{negative}? 
\begin{eqnarray} \label{eq:b-1} 
p_{S+} &=&\Pr \, (\mathcal{H} \, | \, \mathcal{T}) 
\\ [2mm] \label{eq:b-2} 
p_{S-} &=&\Pr \, (\mathcal{H} \, | \, \bar{\mathcal{T}}) 
\\ [2mm] \label{eq:b-3} 
p_{B+} &=&\Pr \, (\bar{\mathcal{H}} \, | \, \mathcal{T}) 
\, = \, 1 - p_{S+} 
\\ [2mm] \label{eq:b-4} 
p_{B-} &=&\Pr \, (\bar{\mathcal{H}} \, | \, \bar{\mathcal{T}}) 
\, = \, 1 - p_{S-} 
\end{eqnarray} 
\indent 
Focusing on Eq.~(\ref{eq:b-1}), above question may be formulated as: 
when applying the test on a particular event 
$\evt_i \in \setD$ of the \textit{real data sample}, 
which is the conditional \textit{posterior probability} 
$p_{S+}$ of $\evt_i$ being signal ($\mathcal{H}$ = \textit{true}) 
\textit{if} the test result is positive ($\mathcal{T}$ = \textit{true})? 
The answer is provided by \textit{Bayes' theorem} \cite{sivia, cern99} 
\begin{equation} \label{eq:b-7} 
\Pr \, (\mathcal{H} \, | \, \mathcal{T}) = 
\frac{\Pr \, (\mathcal{T} \, | \, \mathcal{H})}{\Pr \, (\mathcal{T})} 
\cdot\Pr \, (\mathcal{H}),
\end{equation} 
with the \textit{positive evidence} being calculated by the law of total probability:
\begin{equation} \label{eq:b-8} 
p_{+} \equiv\Pr \, (\mathcal{T}) = 
\Pr \, (\mathcal{T} \, | \, \mathcal{H}) \cdot\Pr \, (\mathcal{H}) + 
\Pr \, (\mathcal{T} \, | \, \bar{\mathcal{H}}) \cdot\Pr \, (\bar{\mathcal{H}}) 
\, = \, \varepsilon \cdot p_S + \beta \cdot p_B,
\end{equation} 
using the variables of 
Eqs. (\ref{eq:x-2}, \ref{eq:s-3}, \ref{eq:s-5}). Hence, from Eqs. (\ref{eq:s-7}, \ref{eq:b-7}) 
\begin{equation} \label{eq:b-9} 
p_{S+} = \frac{\varepsilon}{\varepsilon \cdot p_S + \beta \cdot p_B} \cdot p_S \, = \, 
\frac{p_S}{p_S + (1 - p_S) \, / \, \ell_{+}}.
\end{equation} 
Clearly the posterior probability $p_{S+}$ depends only on the prior probability 
$p_S$ and on the \textit{positive likelihood ratio} $\ell_{+} = \varepsilon / \beta$. 
Fig. \ref{fig:bayes} plots $p_{S+} ( \, p_S )$ for $0\leq\beta / \varepsilon\leq1$. 

Similarly, Eq.~(\ref{eq:b-2}) asks: when applying the test, 
which is the conditional \textit{posterior probability} 
$p_{S-}$ of $\evt_i$ being signal ($\mathcal{H}$ = \textit{true}) 
\textit{if} the test result is negative ($\mathcal{T}$ = \textit{false})? 
Applying \textit{Bayes' theorem} to this case yields:
\begin{equation} \label{eq:b-10} 
\Pr \, (\mathcal{H} \, | \, \bar{\mathcal{T}}) = 
\frac{\Pr \, (\bar{\mathcal{T}} \, | \, \mathcal{H})}{\Pr \, (\bar{\mathcal{T}})} 
\cdot\Pr \, (\mathcal{H}), \qquad 
\mathrm{and \; the} \; \textit{negative  evidence} 
\end{equation} 
\vspace{-4mm} 
\begin{equation} \label{eq:b-11} 
p_{-} \equiv\Pr \, (\bar{\mathcal{T}}) = 
\Pr \, (\bar{\mathcal{T}} \, | \, \mathcal{H}) \cdot\Pr \, (\mathcal{H}) + 
\Pr \, (\bar{\mathcal{T}} \, | \, \bar{\mathcal{H}}) \cdot\Pr \, (\bar{\mathcal{H}}) 
= \alpha \cdot p_S + \eta \cdot p_B = 1 - p_{+} .
\end{equation} 
Reverting to the variables of 
Eqs. (\ref{eq:x-2}, \ref{eq:s-4}, \ref{eq:s-6}, \ref{eq:s-8}), Eq.~(\ref{eq:b-10}) becomes 
\begin{equation} \label{eq:b-12} 
p_{S-} = \frac{\alpha}{\alpha \cdot p_S + \eta \cdot p_B} \cdot p_S \, = \, 
\frac{p_S}{p_S + (1 - p_S) \, / \, \ell_{-}}.
\end{equation} 
Similarly to above, the posterior probability $p_{S-}$ depends only on the 
prior probability $p_S$ and on the \textit{negative likelihood ratio} $\ell_{-} = \alpha / \eta$. 

Calculation of the remaining conditional \textit{posterior probabilities} 
$p_{B+}$ and $p_{B-}$ of $\evt_i$ being background ($\mathcal{H}$ = \textit{false}) 
\textit{if} the test result is positive ($\mathcal{T}$ = \textit{true}) or negative 
($\mathcal{T}$ = \textit{false}), respectively, is trivial by using 
Eqs. (\ref{eq:b-3}--\ref{eq:b-4}). 

\Aprior{} information are the likelihood ratios $\ell_{+}$ and $\ell_{-}$ 
which are obtained from the Monte Carlo simulations, see Section \ref{sec:simul}. 
The prior probabilities $p_S$ and $p_B$ are the signal and background fractions, 
Eq.~(\ref{eq:x-2}). They are determined by Eq.~(\ref{eq:x-4}) 
if the signal to background ratio $R_{S/B}$ is known; 
if not, estimates of them can be calculated from 
the real data sample by a method presented in Section \ref{sec:prior}. 


\section{Event selection} 
\label{sec:opti} 

In many experiments, notably those performed by particle collisions, 
the real data sample $\setD$ has been stochastically accumulated by two 
\textit{independent} Poisson processes for signal and background events, 
respectively.\footnote{ 
\, In real experiments, the raw data are further affected by triggers of data acquisition, the 
detector acceptance, reconstruction efficiencies, pre-selection for the analysis proper, etc. 
These complex aspects, which are beyond the scope of this paper, are discussed e.g. in 
\cite{cup2000, springer}. 
}
Thus, their numbers $N_S$ and $N_B$ behave as random variables with expectation values 
\begin{equation} \label{eq:o-a} 
\left< N_S \right> = p_S \cdot N, \qquad \left< N_B \right> = p_B \cdot N ,
\end{equation} 
where $p_S$ and $p_B$ are fixed values, determined only by the physics laws 
underlying the production of signal and background events; 
they are a-priori unknown, but their values can be estimated as shown in 
section \ref{sec:prior}. 

The base sample $\setD = \setS \cup \setB$ thus obtained may be regarded, 
from a frequentist's point-of-view, 
as one stochastic draw from a virtual population, 
consisting of many repetitions of the experiment at same conditions. 

The marginal standard deviations and the correlation
\footnote{ 
\, The correlation $\rho \, (N_S, N_B)$ is a characteristic of these distributions, 
hence it is a constant. 
It must not be confused with the \textit{empirical correlation} $r \, (N_S, N_B)$ 
which is a random variable, dependent on the actual draw, 
however with expectation value 
$\left<  \, r \, \right> = \rho \, (N_S, N_B) = 0$. 
As can be shown by simulation 
with an increasing number of draws, 
plots of the $r$-distribution approach the Dirac function $\delta \, (r)$, 
thus illustrating the fact that $r \, (N_S, N_B)$ is a consistent estimator 
of $\rho \, (N_S, N_B)$.}
 of the bivariate distribution of $N_S$ and $N_B$ are as follows:
 \begin{equation} \label{eq:o-b} 
\sigma( N_S ) = \sqrt{p_S \cdot N}, \qquad \sigma( N_B ) = \sqrt{p_B \cdot N}, 
\qquad \rho \, (N_S, N_B ) = 0 .
\end{equation} 
Hence, for the total number $N = N_S + N_B$ 
\begin{equation} \label{eq:o-c} 
\left< N_S + N_B \right> = N, \qquad \sigma( N_S + N_B ) = \sqrt{ N } .
\end{equation} 

However, for dealing with the following selection, $\setD$ is \textit{postulated} to be 
a base sample of fixed albeit unknown sizes $N_S$ of $\setS$ (signal) and 
$N_B$ of $\setB$ (background), and their sum $N = N_S + N_B$ 
being the fixed and known size of $\setD = \setS \cup \setB$. 

A rigorous analysis is deferred to part~II of this study. 

\subsection{Selection procedure} 
\label{sec:opti-1} 

After having determined, with sufficient precision, the characteristic values 
$\left[ \, \varepsilon, \alpha, \beta, \eta \, \right]$ 
(\ref{eq:s-3}\nobreakdash--\ref{eq:s-6}) 
from the Monte Carlo samples $\setSp$ and $\setBp$, 
the same test criteria $\vec{\theta}$ are now applied on the full \textit{real data sample} 
$\setD$, yielding $E_S$ selected events to enter sub-sample $\setSa$ and 
$E_B$ rejected events to enter sub-sample $\setBa$. 

Since $\varepsilon, \beta$ and $\alpha, \eta$ are \textit{probabilities} of selection and 
rejection, respectively, 
$E_S$ and $E_B$ are random variables resulting from two independent 
Bernoulli processes, with expectation values 
\vspace{2mm} 
\begin{equation} \label{eq:o-1} 
\left< E_S \right> = \varepsilon \cdot N_S + \beta \cdot N_B 
\; = \; \mathrm{expected \; no. \; of \; events \; to \; be \; selected}, 
\end{equation} 
\vspace{-10mm} 
\begin{equation} \label{eq:o-2} 
\left< E_B \right> = \alpha \cdot N_S + \eta \cdot N_B 
\; = \; \mathrm{expected \; no. \; of \; events \; to \; be \; rejected} .
\end{equation} 
Hence with Eq.~(\ref{eq:s-9}) :
\begin{equation} \label{eq:o-4} 
\sigma (E_S) = \sigma (E_B) = 
\sqrt{N_S \cdot \varepsilon \cdot \alpha \, + \, N_B \cdot \beta \cdot \eta} .
\end{equation} 

The postulated fixed event numbers ``per draw'', $N_S$ and $N_B$, imply 
maximal anti-correlation of selected and rejected events,\footnote{ 
\, A common mistake is tacitly assuming $\rho \, (E_S, E_B ) = 0$.} 
\begin{equation} \label{eq:o-3} 
\left< E_S \right> + \left< E_B \right> = N_S + N_B \qquad 
\Longrightarrow \qquad \rho \, (E_S, E_B ) = -1 ,
\end{equation} 
and error propagation yields 
$\sigma (E_S + E_B) = \left| \, \sigma( E_S ) - \sigma( E_B) \, \right| = 0$. 


\subsection{Test optimization} 
\label{sec:opti-3} 

Optimizing the test parameters $\vec{\theta}$ requires a compromise between minimizing 
both $\alpha$ (loss of genuine signals) and $\beta$ (contamination by genuine background) 
with regard to the sample $\setSa$ of selected events. 
This can be formalized by defining an appropriate \textit{figure of merit} (FOM) as a 
function $F ( \vec{\theta} )$, thus implicitely affecting the test criteria $\alpha( \vec{\theta} )$ 
and $\beta( \vec{\theta} )$, and tuning~$\vec{\theta}$ with the goal of 
maximizing the FOM fuction $F$. 

One might be tempted to base the FOM on real data, e.g. by using the purity $x_S$, 
Eq.~(\ref{eq:z-e}) of Section \ref{sec:quali}. 
However, this poses the danger of artificially enhancing a possible statistical fluctuation, 
hence it must strictly be avoided \cite{bfactories}. 

Therefore, the FOM should be based on simulated data. 
If the signal to background ratio of real data, $R_{S/B}$ (Eq.~(\ref{eq:x-3})), 
is approximately known and is not too small, and the simulated signal and background 
sample sizes are scaled such that $N'_S / N'_B \approx R_{S/B}$, then 
a simple heuristic FOM ansatz may be (see Eq.~(\ref{eq:s-2})):
\begin{equation} \label{eq:o-5} 
F ( \vec{\theta} ) = \frac{n_{S+}}{\, \sqrt{ N'_{+} } \,} = 
\frac{n_{S+}}{\, \sqrt{ n_{S+} + n_{B+} } \,} \, \rightarrow \, \mathrm{max}. 
\end{equation} 
\indent 
For more sophisticated figures of merit see e.g. \cite{bfactories, punzi}. 


\section{Estimation of $p_S$ and $p_B$} 
\label{sec:prior} 

The fractions $p_S$ and $p_B = 1 - p_S$ 
must be obtained from \textit{independent} prior knowledge. 
If the signal to background ratio (SBR) is known, 
i.e. $R_{S/B} = \left< R_{S/B} \right> \pm \sigma (R_{S/B} )$,
their values are determined by Eq.~(\ref{eq:x-4}), 
and error propagation yields 
\begin{eqnarray} \label{eq:p-0} 
& & \sigma (\pStilde ) = \sigma ( \pBtilde ) = 
\frac{\sigma ( R_{S/B} )}{\, ( 1 + R_{S/B} )^2 \,} \qquad \mathrm{and} \qquad 
\rho \, (\pStilde, \pBtilde ) = -1 .
\end{eqnarray} 
Estimates of $N_S$ and $N_B$ are derived straight-forward from Eq.~(\ref{eq:x-2}). 

In the following we assume that SBR is unknown. 
In this non-trivial case the fractions $p_S$ and $p_B$ are unknown parameters, 
however, their estimation is possible 
by using as prior knowledge the test's characteristic values 
$\left[ \, \varepsilon, \alpha, \beta, \eta \, \right]$, 
applied to the real data sample $\setD$ at hand. 
The method is described in detail below. 

The number of real data events selected and rejected are given by 
$E_S = \left< E_S \right> \pm \sigma ( E_S )$ and 
$E_B = \left< E_B \right> \pm \sigma ( E_B )$, respectively, 
with $E_S + E_B = N$. 
These numbers can be used for calculating estimates of the true number of 
signal and background events, $\NStilde$ and $\NBtilde$, respectively, by 
re-writing Eqs.  (\ref{eq:o-1}--\ref{eq:o-2}) as 
\begin{eqnarray} \label{eq:p-1} 
& & \varepsilon \cdot \NStilde + \beta \cdot \NBtilde = E_S ,
\\ [2mm] \label{eq:p-2} 
& & \alpha \cdot \NStilde + \eta \cdot \NBtilde = E_B ,
\end{eqnarray} 
or in matrix notation (see Eq.~(\ref{eq:s-10})) as 
\begin{equation} 
\begin{array}{l} 
\quad \; \cMat \cdot 
\left( \begin{array}{c} 
\NStilde \\[2mm] 
\NBtilde 
\end{array} \right) \; = \; 
\left( \begin{array}{c} 
E_S \\[2mm] 
E_B 
\end{array} \right), 
\qquad \qquad 
\cMat \; = \; 
\left( \begin{array}{cc} 
\varepsilon & \quad \beta \\ [2mm]
\alpha & \quad \eta 
\end{array} \right) .
\end{array} 
\end{equation} 
Using Eqs. (\ref{eq:s-9}, \ref{eq:s-11}) yields the solutions
\footnote{ 
\, $\det \, ( \cMat ) = \delta = \varepsilon - \beta > 0$ 
(see Eq.~(\ref{eq:s-11})) ensures non-singularity. 
The simpler formulae $\NStilde \approx E_S / \varepsilon$, 
$\NBtilde \approx N - E_S / \varepsilon$, as often used instead, 
are valid only for a negligible contamination rate $\beta \ll 1$.} 
\begin{eqnarray} \label{eq:p-3} 
& & \NStilde = 
\frac{(1 - \beta) \cdot E_S - \beta \cdot E_B}{\delta} 
\, = \, \frac{E_S - \beta \cdot N}{\delta} \; \geq 0 ,
\\ [2mm] \label{eq:p-4} 
& & \NBtilde = 
\frac{\varepsilon \cdot E_B - (1 - \varepsilon) \cdot E_S}{\delta} 
\, \, = \, \frac{\varepsilon \cdot N - E_S}{\delta} \; \ge 0 ,
\end{eqnarray} 
obeying $\NStilde + \NBtilde = N$. 
Error propagation, observing 
$\rho \, (E_S, E_B) = - 1$ (see Eq.~(\ref{eq:o-3})), 
yields the corresponding standard deviations as 
\begin{eqnarray} \label{eq:p-5} 
& & \sigma ( \NStilde ) = 
\frac{(1 - \beta) \cdot \sigma ( E_S ) + \beta \cdot \sigma ( E_B )}{\delta}  
\, = \, \frac{\sigma ( E_S )}{\delta} ,
\\ [2mm] \label{eq:p-6} 
& & \sigma ( \NBtilde ) = 
\frac{\varepsilon \cdot \sigma ( E_B ) + (1 - \varepsilon) \cdot \sigma ( E_S )}{\delta} 
\, = \, \frac{\sigma ( E_B )}{\delta} \, = \, \sigma ( \NStilde ) ,
\end{eqnarray} 
with $\sigma( E_S ) = \sigma( E_B )$ given by Eq.~(\ref{eq:o-4}) and 
approximately setting $N_S \approx \NStilde$ and $N_B \approx \NBtilde$.
$\NStilde + \NBtilde = N$ implies maximal anti-correlation:
\begin{eqnarray} \label{eq:p-7} 
& & \rho \, (\NStilde, \NBtilde ) = -1 ,
\end{eqnarray} 
and error propagation yields 
$\sigma (\NStilde + \NBtilde) = 
| \, \sigma( \NStilde ) - \sigma( \NBtilde) \, | = 0$. 

Note that the values of Eqs. (\ref{eq:p-5}--\ref{eq:p-7}) are derived empirically 
from the observed numbers $E_S$ and $E_B$. 
They must not be confused with Eq.~(\ref{eq:o-b}) which represents characteristics 
of the distributions determined by the data acquisition. 

Estimates of the signal and background fractions, $p_S$ and $p_B = 1 - p_S$, 
are derived straight-forward from Eqs. (\ref{eq:p-3}--\ref{eq:p-4}) 
with Eq.~(\ref{eq:x-2}) as 
\begin{eqnarray} \label{eq:p-8} 
& &\pStilde = \frac{\NStilde}{N} \, = \, 
\frac{E_S / N - \beta}{\delta} \, , \qquad \qquad 
\pBtilde = \frac{\NBtilde}{N} \, = \, 
\frac{\varepsilon - E_S / N}{\delta} ,
\end{eqnarray} 
with $\pStilde + \pBtilde = 1$, implying maximal anti-correlation:
\begin{eqnarray} \label{eq:p-9} 
& & \rho \, (\pStilde, \pBtilde ) = -1 .
\end{eqnarray} 
\indent 
The estimated values of Eq.~(\ref{eq:p-8}) can be used as prior probabilities in the 
Bayesian inference formulae, Eqs. (\ref{eq:b-9}, \ref{eq:b-12}) of Section \ref{sec:bayes}. 
Standard deviations are 
\begin{eqnarray} \label{eq:p-10} 
& &  \sigma (\pStilde ) = \frac{\sigma ( \NStilde )}{N} \, , \qquad \qquad 
\sigma ( \pBtilde ) = \frac{\sigma ( \NBtilde )}{N} \, = \, \sigma (\pStilde ) ,
\end{eqnarray} 
with $\sigma( \NStilde ) = \sigma( \NBtilde )$ given by 
Eqs. (\ref{eq:p-5}--\ref{eq:p-6}). 

Accurate estimates of the true number of signal and background events in the 
real data sample $\setD$, see Eqs. (\ref{eq:p-3}--\ref{eq:p-6}), 
are essential for data analyses based on event counting, 
e.g. measurement of particle decay branching fractions. 


\section{Data quality} 
\label{sec:quali} 

Investigation of the sample $\setSa$, containing $E_S$ \textit{selected real events}, 
requires an accurate knowledge of the expected numbers $\left <E_{S/S} \right>$ of pure 
signal events and $\left <E_{B/S} \right>$ of contaminating background events 
within this sample.\footnote{ 
\, Such an investigation would typically be summarized like ``analysis performed on 
$E_S$ events, $\left <E_{B/S} \right>$ of which are expected to be background''.} 

Their fractions w.r.t. all selected events, \textit{purity} \footnote{ 
\, The term ``purity'' is defined ambigously in the literature.} 
and \textit{fraction of contamination}, 
are relevant quantities for judging the data quality: 
\begin{eqnarray} \label{eq:z-a} 
& & x_S = \frac{\, \left< E_{S/S} \right> \,}{E_S}, \qquad 
x_B = \frac{\, \left< E_{B/S} \right> \,}{E_S} \qquad \mathrm{obeying} \qquad 
x_S + x_B = 1.
\end{eqnarray} 

These fractions $x_S$ and $x_B = 1 - x_S$ are the probabilities of a selected event 
to be signal or background, respectively. 
The number $n_S$ ($n_B$) of signal (background) events within the sample 
$\setSa$ of size $E_S$ is distributed according to the binomial pdf: 
\begin{eqnarray} \label{eq:z-b} 
& &\Pr \, (n_S | x_S) = 
{E_S \choose n_S} \cdot x_S^{n_S} \cdot (1 - x_S)^{E_S - n_S} , 
\quad 0 \leq n_S \leq E_S, \\[2mm] 
\label{eq:z-c} 
& &\Pr \, (n_B | x_B) = 
{E_S \choose n_B} \cdot x_B^{n_B} \cdot (1 - x_B)^{E_S - n_B}, 
\quad 0 \leq n_B \leq E_S. 
\end{eqnarray} 

\subsection{Purity} 
\label{sec:quali-0} 

An estimate of the expected number of pure signal events within the selected sample 
$\setSa$ can be derived from Eqs. (\ref{eq:o-1}, \ref{eq:p-3}), yielding 
\begin{equation} \label{eq:z-d} 
\quad \: \left< E_{S/S} \right> = \varepsilon \cdot N_S \approx \varepsilon \cdot \NStilde = 
\frac{\varepsilon}{\delta} \cdot \left( E_S - \beta \cdot N \right) .
\end{equation} 
\indent 
Its fraction w.r.t. all selected events, the \textit{purity}, is 
\begin{equation} \label{eq:z-e} 
\quad \: x_S = \frac{\, \left< E_{S/S} \right> \,}{E_S} \approx  
\frac{\varepsilon}{\delta} \cdot \left( 1 - \beta \cdot \frac{N}{E_S} \right) .
\end{equation} 

\subsection{Fraction of contamination} 
\label{sec:quali-2} 

An estimate of the expected number of background contamination within 
the selected sample $\setSa$, 
under the assumption of a non-zero contamination rate, i.e. $\beta > 0$, 
can be derived from Eqs. (\ref{eq:o-1}, \ref{eq:p-4}), which yields 
\begin{equation} \label{eq:z-1} 
\quad \: \left< E_{B/S} \right> = \beta \cdot N_B \approx \beta \cdot \NBtilde = 
\frac{\beta}{\delta} \cdot \left( \varepsilon \cdot N - E_S \right). 
\end{equation} 

If knowing the \textit{normalization factor} $1 / r_B$, Eq.~(\ref{eq:s-0}), 
together with the expected number $\left< n_{B+} \right> \approx n_{B+}$ of 
\textit{wrong positives} (simulated background passing the test, see Eq.~(\ref{eq:s-5})), 
above Eq.~(\ref{eq:z-1}) is equivalent to 
\begin{equation} \label{eq:z-2} 
\quad \: \left< E_{B/S} \right> = \frac{\left< n_{B+} \right>}{r_B} \approx \frac{n_{B+}}{r_B} = 
\beta \cdot \frac{N'_B}{r_B} = \frac{\beta \cdot N}{\, 1 + R_{S/B} \,} \, , 
\end{equation} 
with \textit{signal to background ratio} of real data, 
Eqs. (\ref{eq:x-3}, \ref{eq:p-3}--\ref{eq:p-4}) 
\begin{equation} \label{eq:z-3} 
\quad \: R_{S/B} = \frac{N_S}{N_B} \approx \frac{\NStilde}{\NBtilde} = 
\frac{\, E_S - \beta \cdot N \,}{\varepsilon \cdot N - E_S} .
\end{equation}

The \textit{fraction of contamination} w.r.t. all selected events is 
\begin{equation} \label{eq:z-4} 
\quad \: x_B = \frac{\, \left< E_{B/S} \right> \,}{E_S} = 
\frac{\left< n_{B+} \right>}{\, r_B \cdot E_S \,} = \beta \cdot \frac{N_B}{E_S} \approx 
\frac{\beta}{\delta} \cdot \left( \varepsilon \cdot \frac{N}{E_S} - 1 \right) = 1 - x_S .
\end{equation} 

\subsection{Zero ``wrong positives''} 
\label{sec:quali-3} 

If the test criteria are so tight that no simulated background event survives the test, 
i.e. $n_{B+} = 0$ implying $\beta = 0$, then the expectation value 
$\left< n_{B+} \right> \neq n_{B+}$, 
and Eq.~(\ref{eq:z-1}) and the approximate part of Eq.~(\ref{eq:z-2}) are no longer valid. 

In this case, an estimate of the expectation value $\left< E_{B/S} \right>$ 
follows Bayesian reasoning.
Given the sample $\setBp$ of $N'_B$ \textit{simulated background} events, and assuming a 
probability $\beta' > 0$ of such an event to pass the test, then the probability 
$\Pr \, (n_{B+} | \, \beta' )$ of observing $n_{B+}$ events 
passing this test is given by the binomial pdf 
\begin{equation} \label{eq:z-5} 
\quad \:\Pr \, (n_{B+} | \, \beta' ) = 
{N'_B \choose n_{B+}} \cdot {\left( \beta' \right)}^{n_{B+}} \cdot 
\left( 1 - \beta' \right)^{N'_B - n_{B+}}. 
\end{equation} 
\indent 
Now we observe $n_{B+} = 0$ events to have passed the test, i.e. 
$\Pr \, (0 \, | \, \beta' ) = \left( 1 - \beta' \right)^{N'_B}$. This gives the
posterior pdf $f (\beta')\propto\Pr \, (0 \, | \, \beta' )$. 
Normalization to 1 yields: 
\begin{equation} \label{eq:z-6} 
\quad \: f (\beta') = 
\frac{\left( 1 - \beta' \right)^{N'_B}}{\, \int_0^1 \left( 1 - b \right)^{N'_B} \mathrm{d}b \,} = 
(N'_B + 1) \cdot (1 - \beta')^{N'_B}, 
\end{equation} 
with the expectation value
\begin{equation} \label{eq:z-7} 
\quad \: \left< \beta' \right> = \int_0^1 \beta' \cdot f (\beta') \, \mathrm{d}\beta' = 
\frac{1}{\, N'_B + 2 \,} \approx {N'_B}^{-1} > 0 
\end{equation} 
being the probability of a background event to pass the test. 

Regarding sample $\setB$ of $N_B$ \textit{real background} events, Eq.~(\ref{eq:z-1}) 
suggests 
\begin{equation} \label{eq:z-8} 
\quad \: \left< E_{B/S} \right> = \left< \beta' \right> \cdot N_B = 
\frac{N_B}{N'_B} = \frac{1}{r_B} \, , 
\end{equation} 
hence in the sample $\setSa$ of \textit{real selected} events, the expected number 
of background events  is equal to the normalization factor. 
Inserting into Eq.~(\ref{eq:z-2}) yields 
\begin{equation} \label{eq:z-9} 
\quad \: \left< n_{B+} \right> = \left< E_{B/S} \right> \cdot r_B = 1 \, , 
\end{equation} 
hence the posterior expectation value of the number of \textit{simulated background} events 
wrongly passing the test is one, while their observed number $n_{B+} = 0$. 



\vspace*{\baselineskip}
\begin{center} 
{\Large \textbf{Part II: Statistical analysis} } 
\end{center} 

\section{Assumptions and notation}\label{sec:Assumptions} 

Let $\D$ be a data set consisting of $N$ recorded events $\evt_i,\ts{}i=1,\ldots,N$. $\D$ can be partitioned into the subsets $\S$  containing the ``signal''  events and  $\B$ containing the ``background'' events:
\begin{gather}
\S\cup\B=\D,\ts{}\S\cap\B=\emptyset,\ts{}|\S|=\NS,\ts{} |\B|=\NB,\ts{} \NS+\NB=N.
\end{gather}
The respective fractions of ``signal''and  ``background'' events are denoted by:
\begin{gather}
\pS=\NS/N,\ts{}\pB=\NB/N=1-\pS.
\end{gather}  
A binary indicator $I_i\in\{0,1\}$ is attached to each event $_i$ such that 
\begin{gather}I_i=1\iff \evt_i\in\S.\end{gather}
 The actual values of the indicators are unknown \aprior. 

A binary classifier is a function $C$ that computes a function value $c_i=C(\evt_i)\in\{0,1\}$ . If $c_i=1$, event $\evt_i$  is classified as ``signal''; if $c_i=0$, event $\evt_i$ is classified as  ``background''.  Such a classifier cannot be expected to perform perfectly. The probabilities of correct and wrong decisions can be summarized as follows:
\begin{align}
\Pr(c_i=1 \mymid I_i=1)&=\epsi=1-\alpha,\notag\\
\Pr(c_i=0 \mymid I_i=1)&=\alpha,\notag\\
\Pr(c_i=1 \mymid I_i=0)&=\beta,\notag\\
\Pr(c_i=0 \mymid I_i=0)&=\eta=1-\beta,\label{eq0}
\end{align}
where $\alpha\geq0$ is the probability of classifying a ``signal'' event as being  ``background'' 
(type I error), and $\beta\geq0$ is the probability of classifying a  ``background'' event as being ``signal'' (type II error). It is required that $\alpha+\beta<1$, see Eq.~(\ref{eq:s-11}). 
The difference $\epsi-\beta=1-\alpha-\beta > 0$ will be denoted by $\delta$. 

The probabilities $\alpha$ and $\beta$ are unknown \aprior, but can be estimated to arbitrary precision from an independent, sufficiently large  simulated sample $\Dmc$, for which the classification of events into ``signal''and  ``background'' is known, see Section~\ref{sec:simul}. In the following, $\alpha$ and $\beta$ will be considered as known constants.

The respective probabilities $\qS,\qB$ of classifying an arbitrary event as ``signal'' or ``background'' are given by:
\begin{gather}
\bmat\qS\\\qB\emat=\A\bmat\pS\\\pB\emat\!,\ \text{with  } \A=\bmat \epsi & \beta\\\alpha & \eta \emat  \text{and  } \qS+\qB=1. \label{eq5}
\end{gather}

The respective numbers $\ES$ and $\EB$  of events classified by $C$ as ``signal'' and  ``background'' are given by:
\begin{gather}
\ES=\sum_{i=1}^N c_i,\ts{}\EB=N-\ES.
\end{gather}

\section{Bayesian estimation}\label{sec:BayesEst}

Conditional on $\qS$, $\ES$ is distributed according to the binomial distribution \Bi{N,\qS} with the probability function:
\begin{gather}
\Pr(\ES=k\mymid\qS)=\binom{N}{k}{\qS}^k(1-\qS)^{N-k},\quad 0\leq k\leq N.
\end{gather}
The posterior pdf of $\qS$, given an observation $\ES$, can be written down in closed form if the prior pdf $\pi(\qS)$ is taken from the conjugate family of priors, which is the Beta family in this case~\cite{BT}.

If the prior distribution of $\qS$  is \Be{a,b}, the posterior pdf is given by:
\begin{gather}
p(\qS\mymid\ES=k)\propto\qS^{k+a-1}(1-\qS)^{N-k+b-1}\propto\Be{\ES+a,\EB+b}.\label{eq:post}
\end{gather}
The Bayes estimator, i.e., the posterior mean and its variance are equal to:
\begin{gather}
\Erw{\qS\mymid\ES}=\frac{\ES+a}{N+a+b},\ts{}\var{\qS\mymid\ES}=\frac{(\ES+a)(\EB+b)}{(N+a+b)^2(N+a+b+1)},
\end{gather}
where $\Erw{X}$ is the expectation of the random variable $X$ and $\var{X}$ is its variance.

If no prior information on $\qS$ is available, there are three popular ``uninformative'' priors 
\footnote{
\, The maximum-entropy prior and Jeffrey's prior are not really uninformative, as they assign a specific probability to any interval $q_1\leq\qS\leq q_2$ that is contained in $(0,1)$, see~\cite{Stark}.
Neither is Haldane's improper prior, as it can be considered as the limit of \Be{\varepsilon,\varepsilon} for $\varepsilon\longrightarrow0$, thus in the limit assigning the probability of 0.5 to both $\qS=0$ and  $\qS=1$. 
}
in the unit interval:
\vspace{5mm} 
\begin{enumerate}[itemsep=-2pt]
\item the maximum-entropy prior  \Be{1,1}=\Un{0,1}: $\pi(\qS)=1,\ 0\leq\qS\leq1$;
\item Jeffrey's prior \Be{\half,\half}: $\pi(\qS)\propto\qS^{-1/2}(1-\qS)^{-1/2},\ 0<\qS<1$;
\item Haldane's improper prior \Be{0,0}: $\pi(\qS)\propto \qS^{-1}(1-\qS)^{-1},\ 0<\qS<1$.
\end{enumerate}

Prior knowledge about $\qS$ can be incorporated via an informative prior. For example, assume that it is known that $\qS$ is about 0.4, with a standard uncertainty of about 0.1. This information can be encoded by a Beta prior \Be{a,b} with mean 0.4 and variance 0.01:
\begin{gather}
\frac{a}{a+b}=0.4,\ts{}\frac{ab}{(a+b)^2(a+b+1)}=0.01\ts{}\Longrightarrow\ts{}a=9.2,\ts{}b=13.8.
\end{gather}

Credible intervals for $\qS$ can be computed via quantiles of the posterior distribution. A symmetric credible interval $C$ with probability content of $1-\alpha$ is given by $C=\left[c_1,c_2\right]$, where $c_1$ is the $\alpha/2$-quantile of the posterior pdf in Eq.~\ref{eq:post} and $c_2$ is the $(1-\alpha/2)$-quantile. 

An HPD interval $[h_1,h_2]$ can be computed by solving the following system of non-linear equations:
\begin{align}
f(h_2)-f(h_1)&=0,\\
F(h_2)-F(h_1)&=1-\alpha,
\end{align}
where $f(h)=p(h\mymid\ES=k)$ is the posterior pdf in Eq.~\ref{eq:post} and $F(h)$ is its cumulative distribution function. A MATLAB code snippet that solves the system is shown in Fig.~\ref{Fig:matlab}.

\begin{figure}[t]
\small\begin{center}
\begin{tcolorbox}[colback=white]
\begin{verbatim}
% function that computes the left-hand side of the system
betahpd=@(h,a,b,alpha) [betapdf(h(2),a,b)-betapdf(h(1),a,b);...
         betacdf(h(2),a,b)-betacdf(h(1),a,b)-1+alpha];
N=100; % sample size
Es=30; % observed signal events
alpha=0.05; % probability outside the interval 
a=1.5; b=2.3F3; % parameters of Beta prior
A=Es+a; B=N-Es+b; % parameters of Beta posterior
% credible interval, starting point for fsolve
hcred=[betainv(alpha/2,A,B);betainv(1-alpha/2,A,B)];
% solve the system
h=fsolve(@(par) betahpd(par,A,B,alpha),hcred);
% h(1)=lower bound, h(2)=upper bound
\end{verbatim}
\end{tcolorbox}
\end{center}
\vspace*{-\baselineskip}
\caption{MATLAB code snippet that computes the bounds of an HPD interval.}\label{Fig:matlab}
\end{figure}

\subsection{Transformation to $\pS$ and $\NS$}

Solving \refeq{eq5} for $\pS$ yields:
\begin{gather}
\pS=\frac{\qS-\beta}{\delta}.
\end{gather} 
As this is an affine transformation, the posterior mean and variance of $\pS$ and $\NS=N\pS$ are given by:
\begin{gather}
\Erw{\pS\mymid\ES}=\frac{\Erw{\qS\mymid\ES}-\beta}{\delta},\ts{}
\var{\pS\mymid\ES}=\frac{\var{\qS\mymid\ES}}{\delta^2},\\
\Erw{\NS\mymid\ES}=N\Erw{\pS\mymid\ES},\ts{}\var{\NS\mymid\ES}=N^2\var{\pS\mymid\ES}.
\end{gather}
The bounds of credible and HPD intervals are transformed in the same way.

\begin{figure}[t]
\hspace*{-10mm}\scalebox{0.85}{\includegraphics{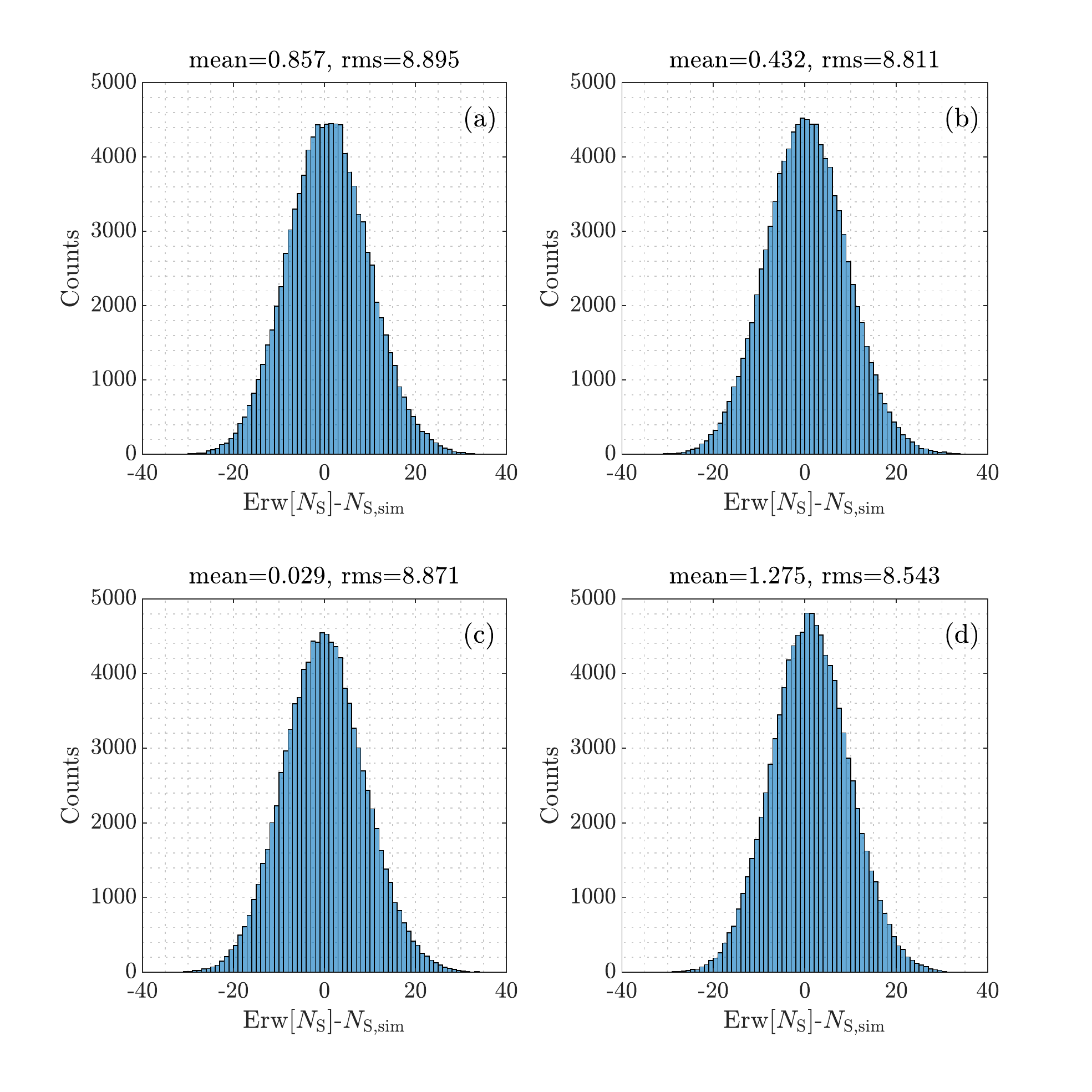}}
\vspace*{-1cm}
\caption{
Histograms of the difference $\Erw{\NS\mymid\ES}-N_\mathrm{S,sim}$ in $M=10^5$ simulated samples with four priors:
(a) Uniform prior, (b) Jeffrey's prior, (c) Haldane's prior, (d) \Be{7.5,42.5} prior. 
}\label{fig:Bayes}
\end{figure}

\paragraph{Example 1} Assume $N=1100, \pS=1/11, \pB=10/11, \alpha=0.1,\beta=0.05$. Then $\epsi=0.9,\eta=0.95,\qS=7/55\approx0.1273,\qB=48/55\approx0.8727.$ Figure~\ref{fig:Bayes} shows histograms of the difference  $\Erw{\NS\mymid\ES}-N_{S,\mathrm{sim}}$ in  $M=10^5$ simulated samples, with three non-informative  priors and an informative prior. The informative prior \Be{7.5,42.5} in subplot (d) encodes the prior information $\qS=0.15$ with a standard uncertainty of 0.05 ($a=7.5,b=42.5$). The estimates in (d) have the largest bias, but the smallest  rms error.

\section{Frequentist estimation}\label{sec:FrequEst}

From a frequentist point of view, the observed data set $\D$ is drawn at random from a (virtual or fictitious) parent population that has to be specified by the person(s) performing the data analysis. From now on, $N$ is a random variable that denotes the size of the elements of the population; the quantities $\NS,\NB,\ES,\EB$ are random variables, too. In contrast to the Bayesian viewpoint, the fraction  $\pS$ of ``signal'' events is an unknown fixed parameters to be estimated. Of course it is assumed that all data sets in the parent population have the same average fraction $\pS$ of ``signal'' events. The size of the observed data set $\D$ is denoted by the constant $\ND$. 

The specification of the parent population is neither unique nor always obvious, and each choice will lead to a different assessment of the unconditional moments of the estimates $\NStilde$ and $\NBtilde$.  However, conditional on the observed size $N$ of the data set, the moments are always the same.
\ifthenelse{\boolean{takesubsecmix}}
{This point is illustrated by three choices of parent populations, the first one with fixed $N$  (Subsection~\ref{subsec:fixed}), the second one with Poisson-distributed~$N$  (Subsection~\ref{subsec:Poisson}),
the third one with $N$ drawn from a mixture of Poisson distributions  (Subsection~\ref{subsec:mixture}).}
{This point is illustrated by two choices of parent populations, one with fixed $N$, the other with Poisson-distributed~$N$.}

\subsection{Data set size $N$ is fixed}\label{subsec:fixed}

If $N$ is fixed, the unconditional distribution of $\NS$ is the same as the conditional distribution, i.e., \Bi{N,\pS}, and the distribution of $\NB$ is \Bi{N,\pB}, with $\pB=1-\pS$. 
It follows that:
\begin{align}
&\Erw{\NS}=N\pS,\Erw{\NB}=N\pB,\\
&\var{\NS}=\var{\NB}=-\cov{\NS}{\NB}=N\pS\pB,\ts{}\corr{\NS}{\NB}=-1,
\end{align}
where \cov{X}{Y} is the covariance of $X$ and $Y$ and $\corr{X}{Y}$ is their correlation. 
It follows that:
\begin{align}
&\Erw{\ES}=N\qS,\ts{}\Erw{\EB}=N\qB,\\
&\var{\ES}=\var{\EB}=-\cov{\ES}{\EB}=N\qS\qB,\ts{}\corr{\ES}{\EB}=-1.\label{ES}
\end{align}
The relation between $\Erw{\NS},\Erw{\NB}$ and $\Erw{\ES},\Erw{\EB}$ can be written in the following way (see \refeq{eq5}):
\begin{gather}
\bmat\Erw{\ES}\\\Erw{\EB}\emat=\A\bmat\Erw{\NS}\\\Erw{\NB}\emat, \text{with  } \A=\bmat \epsi & \beta\\\alpha & \eta \emat.
\label{eq16}
\end{gather}
Using Eq.~(\ref{eq16}), it is easy to show that
\begin{gather}
\bmat\NStilde\\\NBtilde\emat=\H\bmat\ES\\\EB\emat, \text{with  } \H=\A^{-1}=
\frac{1}{\delta}\bmat \eta & -\beta\\-\alpha & \epsi \emat,  \label{eq17}
\end{gather}
is an unbiased estimator of $\NS,\NB$:
\begin{align}
&\Erw{\NStilde}=N\pS,\ts{}\Erw{\NBtilde}=N\pB.\label{eq18}
\end{align}
The joint variance-covariance matrix of $\NStilde,\NBtilde$ can be computed from Eq.~(\ref{eq17}) by linear error propagation:
\begin{align}
&\var{\NStilde}=\var{\NBtilde}=-\cov{\NStilde}{\NBtilde}=N\qS\qB/\delta^2,\corr{\NStilde}{\NBtilde}=-1.\label{eq19}
\end{align}
The relative frequencies $\pS$ and $\pB$ are estimated by:
\begin{gather}
\pStilde=\NStilde/N,\ts{}\pBtilde=\NBtilde/N.
\end{gather}
These estimators are unbiased, and their expectation and \VCM{} is equal to:
\begin{align}
&\Erw{\pStilde}=\pS,\ts{}\Erw{\pBtilde}=\pB,\\
&\var{\pStilde}=\var{\pBtilde}=\qS\qB/(N\delta^2),\\ \label{eqbla}
&\cov{\pStilde}{\pBtilde}=-\qS\qB/(N\delta^2)\ts{}\Longrightarrow\ts{}
\corr{\pStilde}{\pBtilde}=-1.
\end{align}

 $\NStilde$ can also be written as:
\begin{gather}
\NStilde=\frac{\ES-N\beta}{\delta}
\end{gather}
Conditional on $\NS$, the observable $\ES$ is the sum of two random variables $X$ and $Y$, where $X$ ($Y$) is the number of signal (background) events classified as ``signal''. The distribution of $X$ is therefore \Bi{\NS,1-\alpha}, and the distribution of $Y$ is \Bi{\NB,\beta}.
It follows that:
\begin{align}
&\Erw{\ES\mymid\NS}=\NS(1-\alpha)+\NB\beta,\label{eq25}\\ 
&\Erw{\ES-\NS\mymid\NS}=-\NS\alpha+\NB\beta,\\
&\var{\ES\mymid\NS}=\NS\alpha(1-\alpha)+\NB\beta(1-\beta).
\end{align}
Still conditional on $\NS$, we have:
\begin{align}
&\Erw{\NStilde\mymid\NS}=\Erw{\ES/\delta-N\beta/\delta\mymid\NS}\notag\\
&\phantom{\Erw{\NStilde\mymid\NS}}=(1/\delta)\cdot\left(\NS(1-\alpha)+(N-\NS)\beta\right)-N\beta/\delta\notag\\
&\phantom{\Erw{\NStilde\mymid\NS}}=(1/\delta)\cdot\NS\delta+N\beta/\delta-N\beta/\delta=\NS.\\
&\Erw{\NStilde-\NS\mymid\NS}=0\\
&\var{\NStilde\mymid\NS}=(1/\delta^2)\cdot\var{\ES\mymid\NS}=(1/\delta^2)\cdot\left(
\NS\alpha(1-\alpha)+\NB\beta(1-\beta)\right).\label{eq32}
\end{align}
$\NStilde$ and $\NS$ are correlated. From \refeq{eq25} follows:
\begin{align}
\Erw{\ES\NS\mymid\NS}&=\NS^2(1-\alpha)+\NS\NB\beta\folgt\notag\\
\Erw{\ES\NS}&=\Erw{\NS^2}(1-\alpha)+\Erw{\NS(N-\NS)}\beta\notag\\
            &=\delta\Erw{\NS^2}+N\beta\Erw{\NS}\notag\\
            &=\delta\left(N\pS(1-\pS)+N^2\pS^2\right)+N^2\beta\pS\folgt\notag\\
\cov{\ES,\NS}&=\delta N\pS(1-\pS).
\end{align}
As $\NStilde=1/\delta\cdot(\ES-N\beta)$, we finally obtain:
\begin{align}
\cov{\NStilde}{\NS}&=N\pS(1-pS),\label{eq29}\\
\corr{\NStilde}{\NS}&=\delta\frac{\sqrt{\pS(1-\pS)}}{\sqrt{\qS(1-\qS)}}.\label{eq35}
\end{align}
It follows from~\refeq{eq29} that $\NStilde-\NS$ and $\NS$ are uncorrelated:
\begin{gather}
\cov{\NStilde-\NS}{\NS}=\cov{\NStilde}{\NS}-\var{\NS}=0.
\end{gather}

\begin{figure}[t]
\hspace*{-10mm}\scalebox{0.85}{\includegraphics{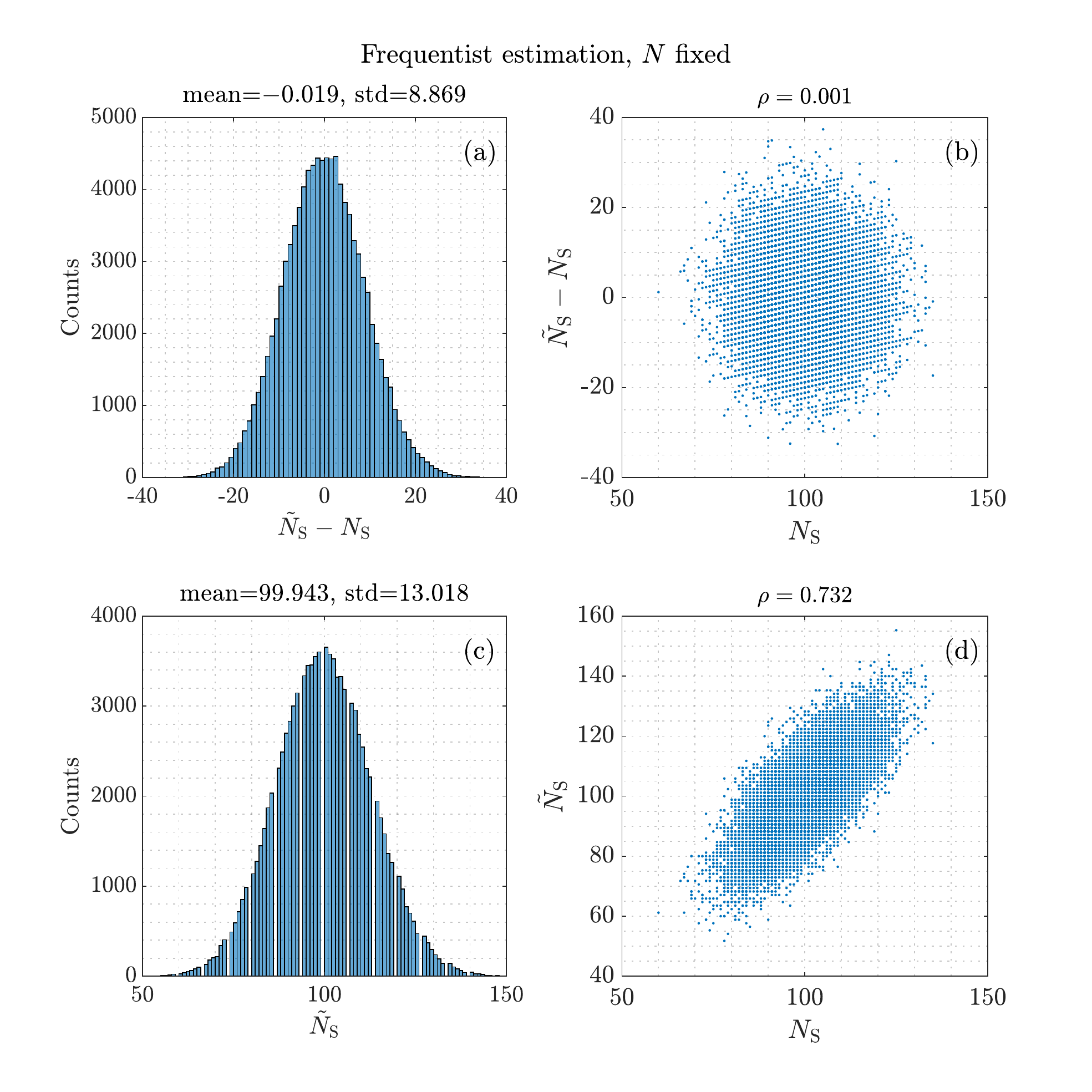}}
\vspace*{-1cm}
\caption{Fixed size $N$ of the data sets. 
(a) Histogram of $\NStilde-\NS$, (b) scatter plot $\NS$ versus $\NStilde-\NS$, (c)~histogram of $\NS$, (d) scatter plot $\NS$ versus $\NStilde$. Note that $\NStilde$ takes only 91 discrete values.
}\label{fig:Freq_11_1}
\end{figure}

\paragraph{Example 2} 
Using the numerical values of Example~1, Table~1 shows the expectations, the standard deviations and the correlation of $\NStilde,\NBtilde$ averaged over a simulated population with $10^5$ data sets. The exact values are shown as well. Figure~\reffig{fig:Freq_11_1} shows histograms and scatter plots of the estimates $\NStilde-\NS$ and $\NStilde$.

\begin{table}[ht]
\newcommand{\myrule}{\rule[-6pt]{0pt}{20pt}}
\newcommand{\mynum}[1]{\makebox[2cm][r]{\ensuremath{#1}}\makebox[1cm]{}}
\caption{Empirical and exact expectations, standard deviations and correlation of $\NStilde,\NBtilde$ with fixed data size $N=1100$.}
$$
\begin{array}{|l||c|c|}
\hline
N=1100 & \makebox[3cm]{{Empirical}} & \makebox[3cm]{{Exact}}\\ \hline
\Erw{\NStilde-\NS}\myrule & \mynum{-0.019} & \mynum{0}\\ \hline
\std{\NStilde-\NS}^{\ (\ast)}\myrule & \mynum{8.87} & \mynum{8.84} \\ \hline
\Erw{\NStilde}\myrule & \mynum{99.94} & \mynum{100.00}\\ \hline
\Erw{\NBtilde}\myrule & \mynum{1000.06} & \mynum{1000.00}\\ \hline
\std{\NStilde}\myrule &  \mynum{13.02} &  \mynum{13.00}\\ \hline
\std{\NBtilde}\myrule &  \mynum{13.02} &  \mynum{13.00}\\ \hline
\corr{\NStilde}{\NBtilde}\myrule  & \mynum{-1.00} &  \mynum{-1.00}\\ \hline
\corr{\NStilde}{\NS}^{\ (\dagger)}\myrule  & \mynum{0.732} &  \mynum{0.733}\\ \hline
\corr{\NStilde-\NS}{\NS}\myrule  & \mynum{0.001} &  \mynum{0}\\ \hline
\end{array}
$$
${}^{\ (\ast)}$ The exact value is the average of the exact standard deviation as given by~\refeq{eq32} over the $10^5$ simulated data sets.\\
${}^{\ (\dagger)}$ The exact value is the average of the exact correlation as given by~\refeq{eq35} over the $10^5$ simulated data sets.
\end{table}

\begin{figure}[t]
\hspace*{-10mm}\scalebox{0.85}{\includegraphics{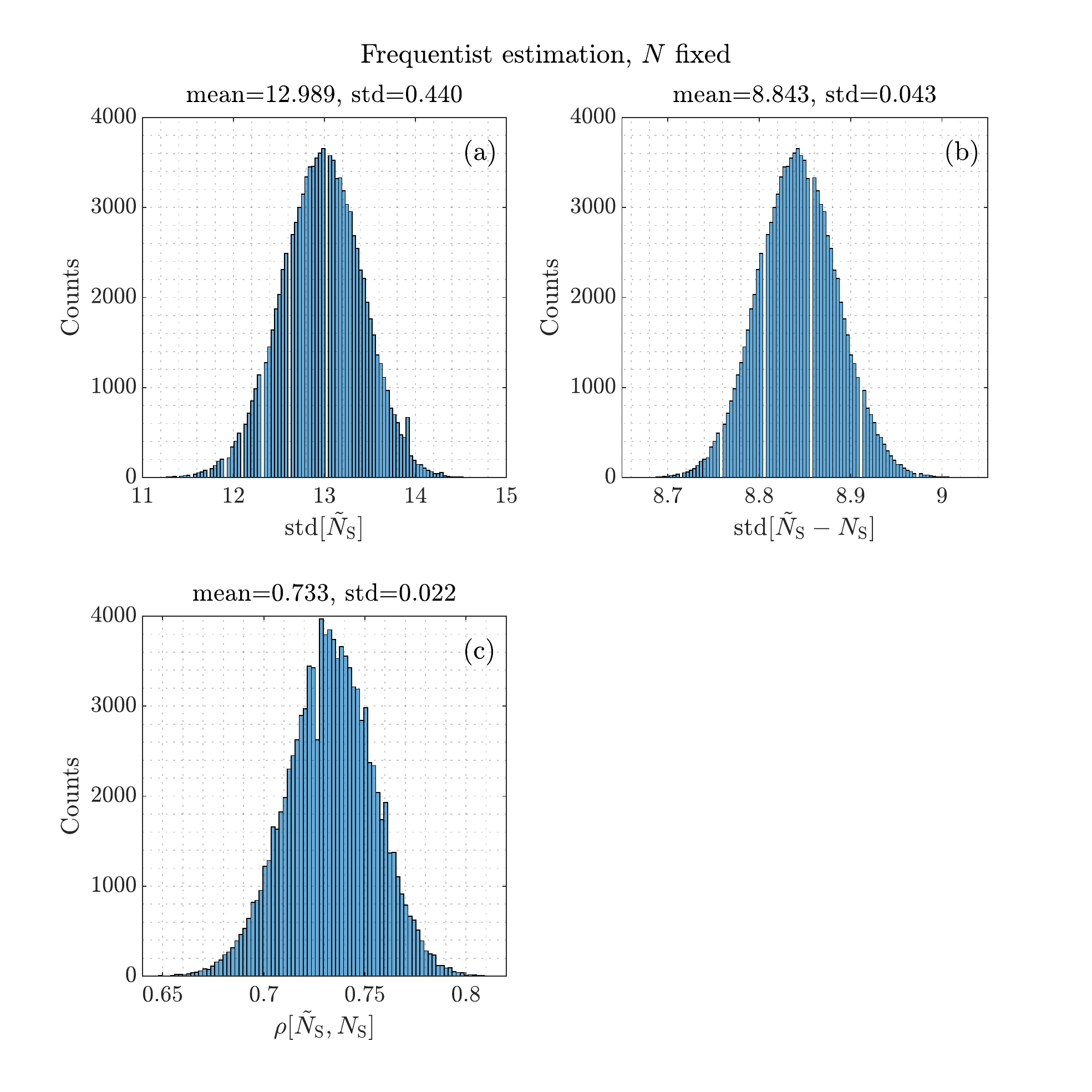}}
\vspace*{-1cm}
\caption{Fixed size $N$ of the data sets. 
Histogram of (a) $\std{\NStilde-\NS}$, (b) $\std{\NStilde-\NS$}, (c) $\corr{\NStilde}{\NS}$. The entries are calculated with the estimated values of $\pStilde$ and $\qStilde$.
}\label{fig:Freq_12_1}
\end{figure}

The variances and covariances derived in this subsection ultimately depend on the unknown parameter $\pS$. In practice, $\pS$ has to be replaced by its estimated value, leading to a deviation from the exact theoretical value. The size of the deviations is illustrated in~\reffig{fig:Freq_12_1}, which shows histograms of $\std{\NStilde}$, $\std{\NStilde-\NS}$ and $\corr{\NStilde}{\NS}$. The entries are the approximative standard deviations that use $\pStilde$ and $\qStilde$ instead of $\pS$ and $\qS$.


\subsection{Data set size $N$ is Poisson-distributed}\label{subsec:Poisson}

Next it is assumed that 
\begin{itemize}
\item each data set in the population is the union of a set $\S$ of size $\NS$ and a set $\B$ of size $\NB$; 
\item $\NS$ and $\NB$ are independent Poisson-distributed random variables with mean values $\lam\pS$ and $\lam\pB$, respectively.  
\end{itemize}
The data set size $N=\NS+\NB$ ist then Poisson distributed according to \Po{\lam}. In the absence of further external information, the obvious choice of the Poisson parameter $\lam$ is $\lam=\ND$.

 Conditional on $N$, we have:
\begin{align}
&\Erw{\ES\mymid N}=N\qS,\ts{}\Erw{\EB\mymid N}=N\qB,\\
&\var{\ES\mymid N}=\var{\EB\mymid N}=N\qS\qB,\\
&\cov{\ES}{\EB\mymid N}=-N\qS\qB,\ts{}\corr{\ES}{\EB\mymid N}=-1.
\end{align}
If $\NS,\NB$ are estimated via Eq.~(\ref{eq17}), their conditional expectation and \VCM{} is given by:
\begin{align}
&\Erw{\NStilde\mymid N}=N\pS,\ts{}\Erw{\NBtilde\mymid N}=N\pB,\\
&\var{\NStilde\mymid N}=\var{\NBtilde\mymid N}=N\qS\qB/\delta^2,\\ \label{eqbla1}
&\cov{\NStilde}{\NBtilde\mymid N}=-N\qS\qB/\delta^2\ts{}\Longrightarrow\ts{}
\corr{\NStilde}{\NBtilde\mymid N}=-1.
\end{align}
The unknown fractions $\pB,\pS$ are estimated by:
\begin{align}
&\pStilde=\NStilde/N,\ts{}\pBtilde=\NBtilde/N.
\end{align}
Conditional on $N$, these estimators are unbiased. Their expectation and conditional \VCM{} is equal to:
\begin{align}
&\Erw{\pStilde\mymid N}=\pS,\ts{}\Erw{\pBtilde\mymid N}=\pB,\\
&\var{\pStilde\mymid N}=\var{\pBtilde\mymid N}=\qS\qB/(N\delta^2),\\ \label{eqbla2}
&\cov{\pStilde}{\pBtilde\mymid N}=-\qS\qB/(N\delta^2)\ts{}\Longrightarrow\ts{}
\corr{\pStilde}{\pBtilde\mymid N}=-1.
\end{align}

The unconditional expectations and elements of the \VCM{} can be calculated by taking the expectation with respect to the distribution of $N$, which is \Po{\lam}. First, the conditional second moments around 0 (i.e., raw moments) are computed:
\begin{align}
&\Erw{\NStilde^2\mymid N}=N\qS\qB/\delta^2+N^2\pS^2,\\
&\Erw{\NBtilde^2\mymid N}=N\qS\qB/\delta^2+N^2\pB^2,\\
&\Erw{\NStilde\NBtilde\mymid N}=N\qS\qB/\delta^2+N^2\pS\pB.
\end{align}
Taking the expectation of $N$ over \Po{\lam} yields:
\begin{align}
&\Erw{\NStilde}=\lam\pS,\ts{}\Erw{\NBtilde}=\lam\pB,\\
&\Erw{\NStilde^2}=\lam\qS\qB/\delta^2+(\lam^2+\lam)\pS^2,\\
&\Erw{\NBtilde^2}=\lam\qS\qB/\delta^2+(\lam^2+\lam)\pB^2,\\
&\Erw{\NStilde\NBtilde}=\lam\qS\qB/\delta^2+(\lam^2+\lam)\pS\pB.
\end{align}
The unconditional \VCM{} is then given by:
\begin{align}
&\var{\NStilde}=\lam(\eta^2\qS+\beta^2\qB)/\delta^2,\\
&\var{\NBtilde}=\lam(\alpha^2\qS + \epsi^2\qB)/\delta^2,\\
&\cov{\NStilde}{\NBtilde}=-\lam(\alpha\eta\qS + \beta\epsi\qB)/\delta^2,\\
&\corr{\NStilde}{\NBtilde}=-\frac{\alpha\eta\qS + \beta\epsi\qB}{\sqrt{(\eta^2\qS+\beta^2\qB)(\alpha^2\qS + \epsi^2\qB)}}.
\end{align}

\begin{table}[t]
\newcommand{\myrule}{\rule[-6pt]{0pt}{20pt}}
\newcommand{\mynum}[1]{\makebox[2cm][r]{\ensuremath{#1}}\makebox[1cm]{}}
\caption{Unconditional empirical and exact expectations, standard deviations and correlation of $\NStilde,\NBtilde$ with data size $N$ distributed according to \Po{\lam} with $\lam=1100$.}
$$
\begin{array}{|l||c|c|}
\hline
\lam=1100 & \makebox[3cm]{{Empirical}} & \makebox[3cm]{{Exact}}\\ \hline
\Erw{\NStilde}\myrule & \mynum{99.94} & \mynum{100.00}\\ \hline
\Erw{\NBtilde}\myrule & \mynum{1000.08} & \mynum{1000.00}\\ \hline
\std{\NStilde}\myrule &  \mynum{13.31} &  \mynum{13.35}\\ \hline
\std{\NBtilde}\myrule &  \mynum{32.92} &  \mynum{32.84}\\ \hline
\corr{\NStilde}{\NBtilde}\myrule  & \mynum{-0.18} &  \mynum{-0.18}\\ \hline
\end{array}
$$
\end{table}

\begin{figure}[t]
\hspace*{-10mm}\scalebox{0.85}{\includegraphics{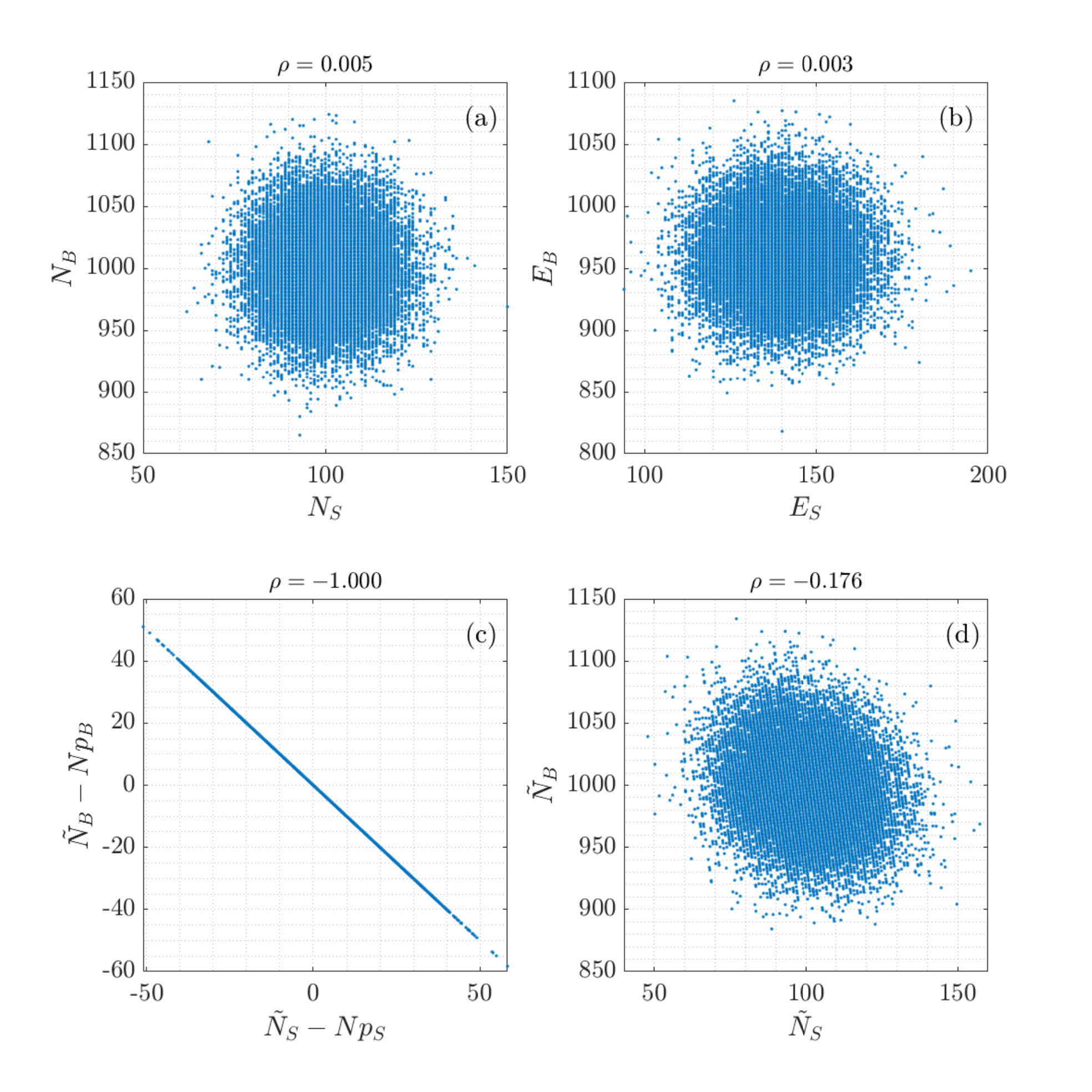}}
\vspace*{-1cm}
\caption{Poisson-distributed  size $N$ of the data sets. Scatter plots of
(a) $\NS$ versus $\NB$, (b) $\ES$ versus $\EB$, (c) $\NS-N\pS$ versus $\NB-N\pB$, (d) $\NStilde$ versus $\NBtilde$. $N$ is distributed according to \Po{\lam} with $\lam=1100$.
}\label{fig:Poptype_2}
\end{figure}

\paragraph{Example 3} 
Using the numerical values of Example~1, the exact conditional expectations, standard deviations and correlation of $\NStilde,\NBtilde$ are identical to the ones shown in Table~1. Table~2 shows the corresponding unconditional values with $\lam=\ND$. \reffig{fig:Poptype_2} shows scatter plots of $\NS$ versus $\NB$, $\ES$ versus $\EB$, $\NS-N\pS$ versus $\NB-N\pB$, and $\NStilde$ versus $\NBtilde$.

%

\ifthenelse{\boolean{takesubsecmix}}{

\subsection{Data set size $N$ is drawn from a mixture\\ of Poisson distributions}\label{subsec:mixture}
An even more general parent distribution is obtained by assuming that the size of the data sets in the population is distributed according to a mixture of Poisson distributions.  This means that the data set size $N$ is drawn from a Poisson distribution with mean $\lam$, and $\lam$ is drawn from a mixing distribution $G$ with mean $\mu$ and variance $\tau^2$. 
The results of the previous subsection show that conditional on $\lam$:
\begin{align}
&\Erw{\ES\mymid\lam}=\lam\qS,\ts{}\Erw{\EB\mymid\lam}=\lam\qB,\notag\\
&\var{\ES\mymid\lam}=\lam\qS,\ts{}\var{\EB\mymid\lam}=\lam\qB,\ts{}\cov{\ES}{\EB\mymid\lam}=0.
\end{align}
This yields the following second raw moments:
\begin{align}
&\Erw{\ES^2\mymid\lam}=\lam^2\qS^2+\lam\qS,\notag\\
&\Erw{\EB^2\mymid\lam}=\lam^2\qB^2+\lam\qB,\notag\\
&\Erw{\ES\EB\mymid\lam}=\Erw{\ES\mymid\lam}\Erw{\EB\mymid\lam}=\lam^2\qS\qB. 
\end{align}
The computation of the unconditional moments relies on the fact that the raw moments of a mixture are a mixture of the raw moments of the mixture components. It follows that:
\begin{align}
&\Erw{\ES}=\ErwG{\lam\qS}=\mu\qS,\notag\\
&\Erw{\EB}=\ErwG{\lam\qB}=\mu\qB,\notag\\
&\Erw{\ES^2}=\ErwG{\lam^2\qS^2+\lam\qS}=(\mu^2+\tau^2)\qS^2+\mu\qS,\\
&\Erw{\EB^2}=\ErwG{\lam^2\qB^2+\lam\qB}=(\mu^2+\tau^2)\qB^2+\mu\qB,\notag\\
&\Erw{\ES\EB}=\ErwG{\lam^2\qS\qB}=(\mu^2+\tau^2)\qS\qB,\notag
\end{align}
where $\ErwG{\cdot}$ denotes the expectation with respect to the mixing distribution $G$. The final step is the computation of the full unconditional variance-covariance matrix of $\ES,\EB$:
\begin{align}
&\var{\ES}=\Erw{\ES^2}-\Erw{\ES}^2=\tau^2\qS^2+\mu\qS,\notag\\
&\var{\EB}=\Erw{\EB^2}-\Erw{\EB}^2=\tau^2\qB^2+\mu\qB,\\
&\cov{\ES}{\EB}=\Erw{\ES\EB}-\Erw{\ES}\Erw{\EB}=\tau^2\qS\qB\notag.
\end{align}
Using Eq.~(\ref{eq17}) and linear error propagation, $\NStilde,\NBtilde$,  their expectations and their joint variance-covariance matrix can be computed, see Subsection~3.2. The explicit formulas are too complicated to be spelled out here. Note that two mixing distributions with the same mean and variance give the same estimates.

The special case where $G$ is a Poisson distribution $\Po{\mu}$ is of particular interest. As $\tau^2=\mu$ in this case, the following variance-covariance matrix of $\ES,\EB$ is obtained:
\begin{align}
&\var{\ES}=\mu(\qS^2+\qS),\ts{}\var{\EB}=\mu(\qB^2+\qB),\ts{}\cov{\ES}{\EB}=\mu\qS\qB.
\end{align}
Linear error propagation gives the following joint variance-covariance matrix of  $\NStilde,\NBtilde$:
\begin{align}
&\var{\NStilde}=\frac{\mu(2\beta^2 - 4\beta\qS + \qS^2 + \qS)}{\delta^2}\notag\\
&\var{\NBtilde}=\frac{\mu(2\alpha^2 - 4\alpha\qB + \qB^2 + \qB)}{\delta^2}\\
&\cov{\NStilde}{\NBtilde}=-\frac{\mu(2\alpha - \qB - 2\alpha\beta - 2\alpha\qB + 2\beta\qB + \qB^2)}{\delta^2}.
\end{align}
In the absence of further external information, the obvious choice of the Poisson parameter $\mu$ is $\mu=\ND$.}{\newpage}

\section{Discussion}\label{sec:Discussion}

Conditional on the data set size $N$, the distribution of the estimates $\NStilde,\NBtilde$ is always the same. The conditional estimates are identical to the Bayes estimate with Haldane's prior, their conditional variance is, however, slightly different from the posterior variance of the Bayes estimator. The unconditional moments of $\NStilde,\NBtilde$ depend of course on the assumptions about the parent population, so different data analysts may come to different conclusions. In addition, if only a single data set has been observed, the conditional moments are much more meaningful. 

An advantage of the Bayesian approach is that prior information about the ``signal'' content of the data set can be easily incorporated into the Bayes estimator by a conjugate Beta prior. On the other hand, with the exception of Haldane's prior, the Bayes estimators have a small bias when compared to the ``truth'' value used in the simulation, and different data analysts may prefer different priors, whether uninformative or informative. 

\vfill

\vfill

\vfill

\vfill



\input{refs.tex} 

\end{document}

%% file: defs.tex
\newcommand{\benum}{\begin{enumerate}[itemsep=2pt,topsep=-2pt]}
\newcommand{\setD}{\mathfrak{D}} 
\newcommand{\setS}{\mathfrak{S}} 
\newcommand{\setB}{\mathfrak{B}} 
\newcommand{\setSa}{\mathfrak{S}^\ast} 
\newcommand{\setBa}{\mathfrak{B}^\ast} 
\newcommand{\setSp}{\mathfrak{S}'} 
\newcommand{\setBp}{\mathfrak{B}'} 
\newcommand{\cMat}{\bm{A}} 
\renewcommand{\Pr}{\mathsf{P}}
\newcommand{\D}{\setD}
\renewcommand{\S}{\setS}
\newcommand{\B}{\setB}
\newcommand{\Dmc}{\setD'}

\newcommand{\A}{\bm{A}}
\renewcommand{\H}{\bm{H}}
\newcommand{\evt}{\mathfrak{e}}
\newcommand{\aprior}{\textit{a priori}}
\newcommand{\Aprior}{\textit{A priori}}

\newcommand{\NS}{N_{S}}
\newcommand{\NB}{N_{B}}
\newcommand{\NStilde}{\tilde{N}_{S}}
\newcommand{\NBtilde}{\tilde{N}_{B}}
\newcommand{\pS}{p_{S}}
\newcommand{\pB}{p_{B}}
\newcommand{\qS}{q_{S}}
\newcommand{\qB}{q_{B}}
\newcommand{\pStilde}{\tilde{p}_{S}}
\newcommand{\pBtilde}{\tilde{p}_{B}}
\newcommand{\qStilde}{\tilde{q}_{S}}

\newcommand{\ES}{E_{S}}
\newcommand{\EB}{E_{B}}
\newcommand{\Bi}[1]{\textsf{Bino}(\ensuremath{#1})}
\newcommand{\Po}[1]{\textsf{Pois}(\ensuremath{#1})}

\newcommand{\Un}[1]{\textsf{Unif}(\ensuremath{#1})}
\newcommand{\Be}[1]{\textsf{Beta}(\ensuremath{#1})}
\newcommand{\Erw}[1]{\langle#1\rangle}
\newcommand{\ErwG}[1]{\Erw{#1}_G}
\renewcommand{\var}[1]{\sigma^2(#1)}
\newcommand{\std}[1]{\sigma(#1)}
\newcommand{\corr}[2]{\rho(#1,#2)}
\newcommand{\cov}[2]{\textsf{cov}(#1,#2)}
\newcommand{\ND}{\ensuremath{N_\setD}}
\newcommand{\bmat}{\begin{pmatrix}}
\newcommand{\emat}{\end{pmatrix}}
\newcommand{\lam}{\lambda}
\newcommand{\refeq}[1]{Eq.~(\ref{#1})}
\newcommand{\reffig}[1]{Fig.~\ref{#1}}
\newcommand{\mymid}{|}
\newcommand{\half}{\ensuremath{\textstyle\frac{1}{2}}}
\newcommand{\bgather}{\begin{gather}}
\newcommand{\egather}{\end{gather}}
\newcommand{\ts}{\;}
\newcommand{\epsi}{\varepsilon}
\newcommand{\VCM}{variance-covariance matrix}
\newboolean{takesubsecmix}
\setboolean{takesubsecmix}{true}
\newcommand{\folgt}{\Longrightarrow}


%% file: refs.tex


%% file: A.bbl
\begin{thebibliography}{99.} 

\bibitem{james}
F. James: 
\textit{Statistical Methods in Experimental Physics,} $2^{nd}$ ed. \\
World Scientific, Singapore 2006 (ISBN 981-270-527-9). 

\bibitem{lyons} 
L. Lyons:
\textit{Statistics for Nuclear and Particle Physicists,} \\
Cambridge University Press, 1992 (ISBN 0-521-37934-2). 

\bibitem{sivia} 
D.S. Sivia: 
\textit{Data Analysis -- A Bayesian Tutorial,} $2^{nd}$ ed. \\
Oxford University Press, 2008 (ISBN 978-0-19-856832-2). 

\bibitem{cern99} 
G. D'Agostini: 
\textit{Bayesian Reasoning in High-Energy Physics}, \\
Yellow Report CERN 99-03, Geneva 1999. 

\bibitem{cup2000} 
R. Fr\"{u}hwirth, M. Regler, R.K. Bock, H. Grote, D. Notz: \\
\textit{Data Analysis Techniques for High-Energy Physics}, $2^{nd}$ ed. \\
 Cambridge University Press, 2000 (ISBN 0-521-63548-9). 

\bibitem{springer} 
C.W. Fabjan and H. Schopper (editors): \\
\textit{Particle Physics Reference Library}, vol. 2. \\
Springer Open, Berlin 2020 
(\href{https://doi.org/10.1007/978-3-030-35318-6}{DOI: 10.1007/978-3-030-35318-6}). 

\bibitem{bfactories} 
F. Porter (editor): 
\textit{The Physics of the B Factories,} chapter 4, \\
Eur.Phys.J. C \textbf{74} (2014) 3026, pp. 59--66, and \\
Springer Open, Berlin 2014 
(\href{https://arxiv.org/abs/1406.6311}{arXiv:1406.6311 [hep-ex]}). 

\bibitem{punzi} 
P. Feichtinger et al.: 
\textit{Punzi-loss}, \\
Eur.Phys.J. C \textbf{82} (2022) 121 
(\href{https://arxiv.org/abs/2110.00810}{arXiv:2110.00810 [hep-ex]}). 

\bibitem{BT}
J.M.~{B}ernardo  and A.F.M.~{S}mith:
\newblock {\em {Bayesian Theory}}.\\
\newblock John Wiley\&Sons, Hoboken NJ, 1994 (ISBN 0-471-92416-4).

\bibitem{Stark}
P.B.~Stark: 
{\textit{Constraints versus Priors}},\\
SIAM/ASA J.~Uncertainty Quantification \textbf{3} (2015) 586 \\
(\href{https://doi.org/10.1137/130920721}{DOI: 10.1137/130920721}).

\end{thebibliography}
